# Extraordinary focusing of sound above a soda can array without time reversal

A A Maznev[1,2,4], Gen Gu[1], Shu-yuan Sun[1], Jun Xu[3], Yong Shen[1], Nicholas Fang[3] and Shu-yi Zhang[1,4]

[1] Laboratory of Modern Acoustics, Institute of Acoustics, Collaborative Innovation Center of Advanced Microstructures, Nanjing University, Nanjing 210093, China

[2] Department of Chemistry, Massachusetts Institute of Technology, Cambridge, Massachusetts 02139, USA

[3] Department of Mechanical Engineering, Massachusetts Institute of Technology, Cambridge, Massachusetts 02139, USA

E-mail: alexei.maznev@gmail.com and zhangsy@nju.edu.cn

**Abstract**

Recently, Lemoult et al. [Phys. Rev. Lett. **107**, 064301 (2011)] used time reversal to focus sound above an array of soda cans into a spot much smaller than the acoustic wavelength in air. In this study, we show that equally sharp focusing can be achieved without time reversal, by arranging transducers around a nearly circular array of soda cans. The size of the focal spot at the center of the array is made progressively smaller as the frequency approaches the Helmholtz resonance frequency of a can from below, and, near the resonance, becomes smaller than the size of a single can. We show that the locally resonant metamaterial formed by soda cans supports a guided wave at frequencies below the Helmholtz resonance frequency. The small focal spot results from a small wavelength of this guided wave near the resonance in combination with a near field effect making the acoustic field concentrate at the opening of a can. The focusing is achieved with propagating rather than evanescent waves. No sub-diffraction-limited focusing is observed if the diffraction limit is defined with respect to the wavelength of the guided mode in the metamaterial medium rather than the wavelength of the bulk wave in air.

[4] Corresponding authors

## 1. Introduction

Diffraction limits the imaging resolution in acoustics and optics to about half a wavelength. In 2001, Pendry [1] showed that a slab of an idealized negative index material would produce perfect images with resolution not limited by the wavelength, which would be achieved by focusing evanescent rather than propagating waves. Even though this idea has been shown not to work in practice for far-field imaging [2], it has stimulated an active search of ways to overcome the diffraction limit [3-7]. One strategy, proposed by Lemoult et al. [8-11] for both electromagnetic and acoustic waves, relies on the time-reversal focusing in a locally resonant metamaterial. In acoustics, this approach has been implemented in a simple and neat experiment with an array of ordinary soda cans serving as Helmholtz resonators [9]. Broad band sound with a center wavelength of about 0.8 m was focused onto a single can with the focal spot size of about 1/25 of the wavelength in air.

In the present study, we aim to answer two questions: (i) Is time reversal essential for achieving the extraordinary focusing demonstrated in Ref. [9]? (ii) Does the observed effect truly beat the diffraction limit with respect to the wavelength of the acoustic wave propagating in the metamaterial medium formed by soda cans? To answer the first question, we arrange soda cans into a nearly circular array and focus monochromatic sound into the center of the array. The focal spot becomes progressively smaller as the acoustic frequency approaches the Helmholtz resonance of the cans from below, with the width of the intensity distribution getting as narrow as 1/40 of the wavelength in air. We demonstrate, based on an analytical effective-medium model as well as finite-element calculations that the observed effect is due to a guided acoustic wave supported by the soda can array at frequencies below the Helmholtz resonance. On approach to the resonance, the wavelength of the guided mode becomes smaller and the degree of its confinement to the array increases. We argue that the sharp focusing, albeit impressive if compared to the wavelength in air, does not beat the diffraction limit with respect to the wavelength of the guided mode propagating in the system.

## 2. Experiment

The experimental arrangement is shown in figure 1. 37 empty soda cans were arranged in a hexagonal array, with 6 commercial speakers placed symmetrically around the array. We used Coka-Cola cans having a volume of 350 cm$^3$ (nominal beverage volume 330 ml) and an opening area of 4 cm$^2$ similar to those used by Lemoult et al. [9], with the fundamental Helmoltz resonance at 420 Hz. A microphone was suspended at a height of 12±2 mm above the top of the cans. A mechanical delay line with 40 cm travel range was used to move the microphone horizontally along a diameter of the array as shown in figure 1(b). The speakers were driven continuously at a given frequency while the microphone was measuring a spatial profile of the acoustic intensity with 0.61 cm steps.

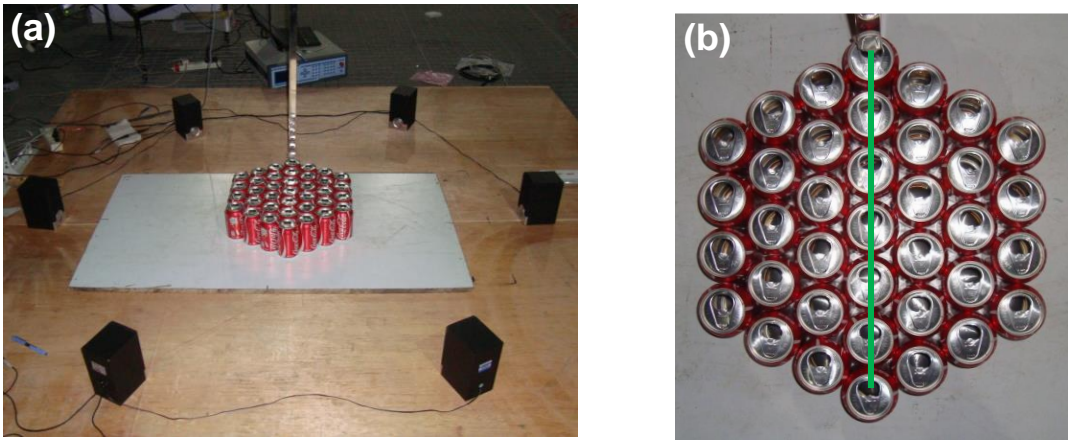

Figure 1. (a) Experimental arrangement inside the anechoic room; (b) top view of the array with the scan line shown.

Figure 2 shows acoustic intensity profiles at different frequencies as well as a reference profile measured without cans. The latter yields a focal spot with a FWHM (full width at half maximum) of 31 cm at 410 Hz, which amounts to 0.37 of the wavelength. The amplitude of a converging cylindrical wave is described by a Bessel function J$_0$(kr) where $k$ is the wavevector and $r$ the distance from the focal point, which yields a FWHM of 0.36λ for the intensity, in a good agreement with the experiment considering that the speakers do not produce a perfect cylindrical wave. In the measurements with soda cans, the focal spot at the center of the array gets progressively smaller as the resonant frequency is approached from below, becoming as narrow as 2 cm, or about λ/40, at 415 Hz. Above the Helmholtz resonance frequency, the intensity profile changes dramatically with maxima at the edges of the array and attenuation towards the center.

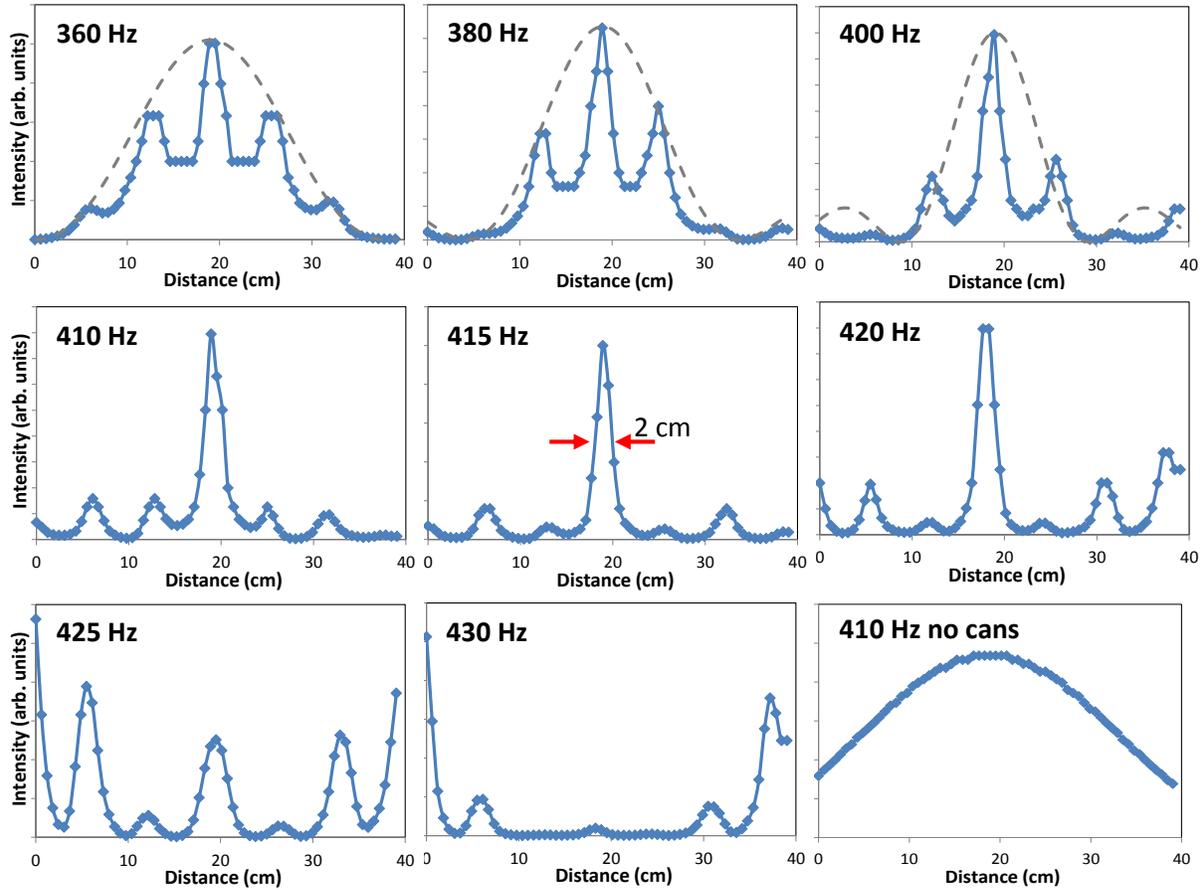

Figure 2. Acoustic intensity profiles along the diameter of the array at different frequencies. The bottom right panel shows an intensity profile measured without soda cans. Symbols are experimental points, connecting lines are guides to the eye. Dashed lines are Bessel function envelopes as per equation (11).

One may ask whether the effect we see is focusing or the excitation of an eigenmode of the array which may just happen to have a sharp maximum at the center. Indeed, a narrow peak in the intensity profile does not necessarily indicate focusing. For example, measurement on a single can near the resonance frequency also yields a sharp peak above the opening of the can, as can be seen in figure 3(a). To verify that we do in fact focus acoustic waves at the center of the array, we changed the arrangement of the speakers by moving all of them by 20 cm along the scan axis as shown in figure 3(c). As one can see in figure 3(b), the intensity maximum shifts by one can in the same direction. In contrast, in the case of a single can the intensity peak remains at the same position as one can see in figure 3(a). That a large displacement of the speakers is required to produce a small displacement of the intensity maximum is due to the fact that the wavelength of the acoustic mode guided by the array is smaller than that in air, as we will see in the next section.

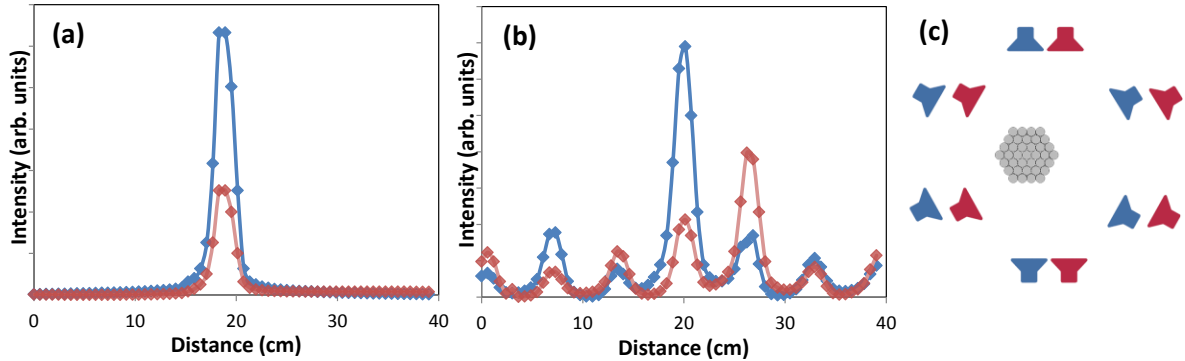

Figure 3. Acoustic intensity profiles at 410 Hz above (a) a single can and (b) the array, with the speakers centered with respect to the can/array (blue) and shifted by 20 cm in the positive scan direction (red). Symbols are experimental points, connecting lines are guides to the eye. (c) Positions of the speakers in the centered (blue) and shifted (red) configurations.

Just as a solid sphere can focus a convergent spherical wave without aberrations only to the center of the sphere, our array yields the best focusing of a cylindrical wave to the center can; focusing is expected to deteriorate as the focal point moves away from the center. Focusing on any can in the array would require shaping the incident field as has been done, for example, in the time reversal method employed in Ref. [9].

## 3. Theory: waveguiding by 2D locally resonant metamaterial

Before discussing the origin of the observed focusing phenomenon, we need to elucidate the nature of waves propagating in the metamaterial formed by the soda cans. It is well known that in a medium containing resonant inclusions, propagating waves hybridize with the local resonance resulting in an avoided crossing bandgap. In optics, this phenomenon forms the basis of the classical dispersion theory (Lorentz oscillator model) [12]. In acoustics, it has attracted renewed interest more recently [13-19] in the context of sound propagation in artificial media that came to be referred to as locally resonant metamaterials [13-15]. One well known example of such locally resonant medium, considered theoretically even before the advent of the metamaterials era, is an array of Helmholtz resonators in a duct exhibiting an avoided crossing bandgap at the Helmholtz resonance [16, 19-22]. The peculiarity of our case is that we have an acoustic wave in a 3D medium interacting with a 2D array of resonant inclusions. In this case, the behavior is different from the classic "avoided crossing" picture described above. As will be shown below both with a simple effective medium model and finite element analysis, an infinite 2D array of resonators supports a guided mode that only exists below the resonant frequency. On the approach to the resonance, this mode becomes increasingly confined while its phase velocity and wavelength become progressively smaller.

*3.1. Effective medium model*

In the effective medium approach, the acoustic wavelength is assumed to be much greater than the average distance between the resonators [14]. The resonators are modeled as mass-on-a-spring harmonic oscillators with pistons of mass $M$ attached to springs with spring constant $K$ as shown in figure 4. The resonators are randomly or regularly distributed in two dimensions, with the average fractional piston area $F$.

For a Helmholtz resonator with a zero neck length, the effective mass is estimated as $(16/3)\pi^{-3/2}\rho_0 A^{3/2}$, where $A$ is the opening area and $\rho_0$ is the density of air, whereas the spring constant $K$ is given by $\rho_0 c^2 A^2/V$, where $c$ is the speed of sound and $V$ is the volume of the resonator [22]. For our soda cans, this model yields an effective mass of 9.23 mg and a spring constant of 64.8 N/m, resulting in a resonance frequency $\omega_0/2\pi = 422$ Hz in good agreement with the experiment [9].

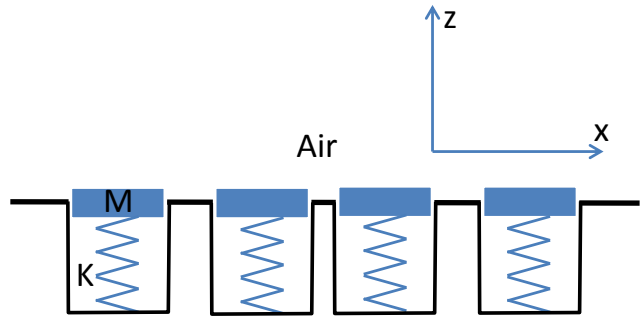

Figure 4. The model system.

The vertical position of a piston $Z$ obeys the equation of motion

$$\ddot{Z} = -\omega_0^2 Z - \frac{pA}{M}, \qquad (1)$$

where $p$ is the deviation of the pressure above the piston from the equilibrium value and $A$ is the area of the piston. If the wavelength of sound is large compared to the average distance between the resonators, the sound wave "senses" an average displacement of the boundary $u_z = FZ$. This leads to an effective boundary condition relating the average displacement and pressure at the boundary z=0,

$$\ddot{u}_z + \omega_0^2 u_z = -\frac{pFA}{M}, \tag{2}$$

which replaces the boundary condition of zero displacement at the rigid boundary in the absence of resonators.

The acoustic velocity potential in the half-space above the resonators is described by the wave equation,

$$\frac{\partial^2 \varphi}{\partial t^2} - c^2 \Delta \varphi = 0. \tag{3}$$

We are looking for a harmonic wave propagating along the floor in the $x$ direction, $\varphi = \tilde{\varphi}(z)\exp(i\omega t - ikx)$, which yields the following equation for the $z$-dependence,

$$\frac{\partial^2 \tilde{\varphi}}{\partial z^2} = \left( k^2 - \frac{\omega^2}{c^2} \right) \tilde{\varphi}, \tag{4}$$

where $c$ is the speed of sound in air. The general solution is given by

$$\tilde{\varphi} = ae^{-\gamma z} + be^{\gamma z}, \tag{5}$$

where

$$\gamma = \left( k^2 - \frac{\omega^2}{c^2} \right)^{1/2}. \tag{6}$$

At $\omega > ck$, we get an imaginary $\gamma$ yielding a bulk wave propagating at an oblique angle; such solutions are of no interest to us. At $\omega < ck$, we get a real $\gamma$, in which case coefficient $b$ in equation (5) should be equated to zero to eliminate the unphysical divergent term, and the potential is given by

$$\varphi = ae^{-\gamma z} e^{i\omega t - ikx}, \tag{7}$$

which describes a guided wave propagating along $x$ and exponentially decaying along z. Plugging equation (7) into the boundary condition given by equation (2) by expressing the displacement and pressure in terms of the potential,

$$\frac{\partial u_z}{\partial t} = \frac{\partial \varphi}{\partial z}; \quad p = -\rho_0 \frac{\partial \varphi}{\partial t}, \tag{8}$$

we arrive to a dispersion relation for $\omega$ and $k$,

$$\left(k^2 - \frac{\omega^2}{c^2}\right)^{1/2} (\omega_0^2 - \omega^2) = \frac{\omega^2 FA}{M}. \tag{9}$$

Figure 5 shows the frequency dispersion according to equation (9) as well as the behavior of the confinement length $1/\gamma$ for $\rho_0$=1.23 kg/m$^3$, $c$=343 m/s, $F$=0.106 (calculated for dense hexagonal packing of the cans), with the resonator parameters listed earlier in the text. In the limit of small $k$ the dispersion approaches that of the bulk wave in air, $\omega = ck$, whereas in the opposite limit of large $k$ the frequency asymptotically approaches the resonance frequency $\omega_0$. On approaching the resonance the guided mode becomes increasingly confined to the floor, as one can see in figure 5(b). The dispersion curve does resemble the lower dispersion branch of the classic "avoided crossing" case [19]. However, there is no upper branch: above $\omega_0$, the frequency acquires an imaginary part, or, for a real frequency (as is the case in our experiment), the wavevector acquires an imaginary part, and the mode becomes evanescent. It should be noted that a bulk wave with a wavevector along $x$, which can propagate along a rigid floor, no longer satisfies the boundary condition. The only wave propagating along the array is the guided mode at frequencies below $\omega_0$.[5] Thus the situation is principally different from the conventional avoided crossing behavior of an array of resonators in a duct [16, 19-21].

According to equation (9), on approaching the resonance the wavevector increases indefinitely, hence the wavelength becomes infinitely small. However, the effective medium model is only valid as long as the wavelength is much greater than the distance between the cans. For a more accurate description of the wave propagation above a hexagonal lattice of soda cans, we use finite element (FE) calculations.

---

[5] Experimental results shown in figure 1(c) of Ref. [9] also indicate that only frequencies below the Helmholtz resonance can propagate along the array.

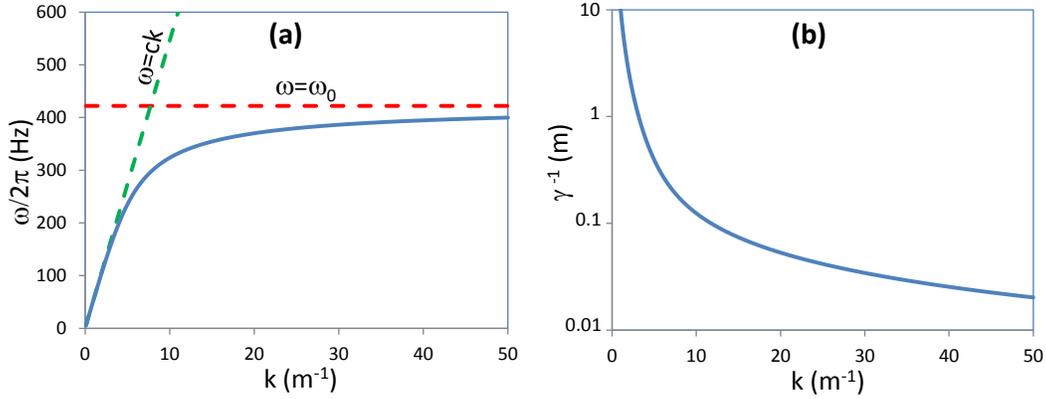

Figure 5. (a) Frequency of the guided wave (solid line) and (b) its confinement length as functions of the wavevector.

*3.2. Finite element modeling*

We used the acoustic module in COMSOL Multiphysics to calculate the dispersion relation of soda cans arranged in a hexagonal lattice. Soda cans were modeled as cylinders with rigid walls [9] of 11.5 cm in height and 6.6 cm in diameter. The opening of a can in the model was circular and centered at the axis of the cylinder, with the same area of 4 cm$^2$ as the opening of a real can. The height of the simulation domain was 1 m, with rigid wall boundary conditions at the "ceiling". Floquet periodic boundary conditions were applied in order to find acoustic eigenmodes of an infinite 2D hexagonal lattice.

Figure 6(a) shows the calculated dispersion along the ΓK direction of the reciprocal lattice. Discrete modes in the shaded area above the sound line $\omega = ck$ are due to a finite height of the simulation domain limited by the computational resource. For a semi-infinite half space, the shaded area will be filled by a continuum of bulk waves propagating at oblique angles to the floor. The mode below the sound line is guided by the can array, and its dispersion is close to what the simple effective medium theory has predicted. Figure 6(b) shows the distribution of the sound pressure in the guided mode in the vertical cross section above a soda can at frequencies 360 and 420 Hz. While the effective medium model correctly predicts increased confinement of the guided mode to the can array on approaching the resonance, FE calculations show that the acoustic field becomes localized at the opening of a can, in agreement with the experiment.

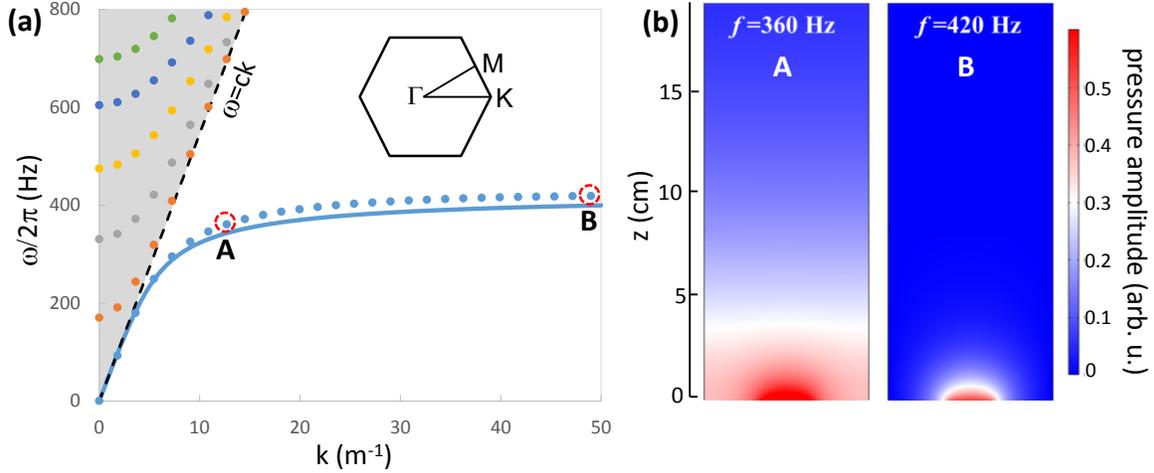

Figure 6. (a) Dispersion of acoustic waves propagating above the hexagonal array of soda cans calculated by FE (symbols) along the ΓK direction of the reciprocal lattice (see the inset) vs. the effective medium calculation (solid line). Shaded area represents the continuum of bulk modes in the semi-infinite space whereas FE calculations yield discrete modes due to the finite height of the simulation domain. (b) Distributions of the sound pressure amplitude in the guided mode above a soda can for points A and B of the dispersion curve shown in (a).

## 4. Discussion

As we have seen, the wave propagating above a 2D array of soda cans is a guided mode whose wavelength gets progressively shorter on approaching the Helmholtz resonance of the cans. This is the principal reason for the sharp focusing observed both in our experiment and in Ref. [9]. The metamaterial structure acts similarly to a solid immersion lens made of a material with a high refractive index to enhance the imaging resolution [23]. The time-reversal technique used in Ref. [9] certainly has an advantage in that it allows one to shape the incident acoustic field to focus on any can and does not require the can array to be shaped as an immersion lens. However, time-reversal is not what makes the intensity maximum at the focal point so narrow.

The focusing is achieved with propagating rather than evanescent waves. Even though the guided mode decays along the $z$ direction, this decay has no bearing on the focusing in the $x$-$y$ plane, in which the wave is propagating. Thus the observed phenomenon has no relation to the imaging by evanescent waves described by Pendry [1].

Superimposed on the focusing phenomenon is the near-field structure of the acoustic field with sharp maxima above the openings of the cans. It is instructive to discuss this behavior in terms of the Bloch expansion of eigenmodes of a periodic structure,

$$\varphi = e^{i\mathbf{k}\cdot\mathbf{r}} \sum_{\mathbf{G}} c_{\mathbf{G}} e^{i\mathbf{G}\cdot\mathbf{r}}, \qquad (10)$$

where $\mathbf{r}=(x,y)$, $\mathbf{k}$ is the reduced wavevector, and $\mathbf{G}$ is a reciprocal lattice vector, with the dependence on $z$ ignored for the sake of simplicity. The sum in equation (10) is a periodic function of $\mathbf{r}$ reflecting the periodicity of the lattice. In the low frequency limit the acoustic dispersion in a hexagonal lattice is isotropic, and the periodic term in equation (10) is nearly independent on the direction of the reduced wavevector. Consequently, we can construct a cylindrical wave described by a Bessel function just as in the isotropic case, with the intensity profile given by

$$|\varphi|^2 = J_0^2(kr)\,P(\mathbf{r}), \qquad (11)$$

where P($\mathbf{r}$) is a lattice-periodic function. Thus the Bessel function will yield the envelope of the profile with a FWHM of ~0.36$\lambda$, whereas the periodic term will define the fine structure of the field. This model is not expected to be accurate for a finite structure used in our experiment (as opposed to an infinite 2D array) because of reflections from the edges of the structure. However, it provides a qualitatively correct description of the behavior seen in figure 2 for frequencies 360-400 Hz, where the Bessel function envelopes calculated according to equation (11), with wavevector values from the FE calculations, are shown by dashed lines. When the ratio of the central peak height to that of its neighbors exceeds a factor of two, as exemplified by the 400 Hz profile in figure 2, the FWHM of the intensity profile will be determined by the central peak width alone, which, close to the resonance, is determined by the size of the opening of a can.

Thus the combination of the reduced wavelength and the fine periodic structure of the acoustic field makes the "focal spot" appear extra small, with a width as narrow as 2 cm which amounts to less than 1/3 of the lattice constant. However, as pointed out in Ref. [9], since one cannot focus between the cans, the actual "resolution" of the focusing does not exceed a single lattice constant. Furthermore, if we apply the traditional definition of the imaging resolution as the ability to distinguish two separate objects from a single object [12] and think of a single can as representing a "pixel" of the object, then the best achievable resolution would be limited by twice the lattice constant. Since on approaching the resonance the acoustic wavelength also becomes as small as about twice the lattice constant (which corresponds to the Brillouin zone boundary in the wavevector space), there is no "sub-wavelength" resolution. The appearance of the latter results from a comparison with the acoustic wavelength in air rather than with the wavelength of the guided acoustic wave propagating in the structure.

We realize that defining the focal spot size, resolution and diffraction limit in complex media is wrought with difficulties. In a periodic medium the wavevector is not uniquely defined; rather, it is defined modulo **G** [24]. Consequently, the wavelength is not uniquely defined either. It is common to define the wavelength based on the reduced wavevector within the first Brillouin zone, but this imposes a low bound on the wavelength, which appears to be unphysical if we consider the limit of vanishing periodicity. What is then the diffraction limit in a periodic medium? We do not have a ready answer and hope that this report will stimulate a discussion of this difficult issue. However, within the metamaterial (i.e., the effective medium) model the picture is clear: focusing gets sharper as the wavelength of the mode guided by the locally resonant metamaterial gets shorter on approach to the resonance.

## 5. Conclusions

We have demonstrated that focusing of sound in a metamaterial formed by a 2D array of soda cans results in an increasingly narrow intensity peak as the acoustic frequency approaches the Helmholtz resonance from below. We conclude that the broad-band time-reversal technique [9], while possessing remarkable capabilities in manipulating sound fields, is not an enabling factor in achieving the sharp focusing. The observed phenomenon results from the small acoustic wavelength in the metamaterial in combination with a near-field effect, i.e., the localization of the acoustic intensity at the opening of a can at frequencies close to the resonance. Even though the intensity peak as narrow as 1/40 of the acoustic wavelength in air has been observed, we do not believe that this results, if properly interpreted, violates the diffraction limit. Furthermore, we found that the acoustic wave propagating along the Helmholtz resonator array is a guided mode becoming increasingly confined to the array as its frequency approaches the Helmholtz resonance from below. In contrast to the well documented case of an "avoided crossing" bandgap, there is no upper branch above the resonance frequency. The phenomenon of "locally resonant" waveguiding will be encountered in other physical systems; indeed, the dispersion of the lowest mode electromagnetic wave guided by an array of metal wires [10,25] is similar to the dispersion of the sound wave guided by soda cans.

**Acknowledgments**

The work performed at Nanjing University was supported by the National Natural Science Foundation of China, Grants No. 11174142, No. 113041600, and No. 11404147, as well as by the National Basic Research Program of China, Grant No. 2012CB921504; the work performed at MIT was supported by the Defense Threat Reduction Agency, Grant No. HDTRA 1-12-1-0008. J. X. and N. F. also acknowledge financial support by Multidisciplinary University Research Initiative from the Office of Naval Research, Grant N00014-13-1-0631.

**References**

[1] Pendry J B 2000 *Phys. Rev. Lett.* **85** 3966

[2] Podolskiy V A and Narimanov E 2005 *Opt. Lett.* **30** 75

[3] Jacob Z, Alekseyev L V and Narimanov E 2006 *Opt. Express* **14** 8247

[4] Liu Z, Lee H, Xiong Y, Sun C and Zhang X 2007 *Science* **315** 1686

[5] Lerosey G, de Rosny J, Tourin A and Fink M 2007 *Science* **315** 1120

[6] Leonhardt U 2009 *New J. Phys.* **11** 093040

[7] Sukhovich A, Merheb B, Muralidharan K, Vasseur J O, Pennec Y, Deymier P A and Page J H 2009 *Phys. Rev. Lett.* **102** 154301

[8] Lemoult F, Lerosey G, de Rosny J and Fink M 2010 *Phys. Rev. Lett.* **104** 203901

[9] Lemoult F, Fink M and Lerosey G 2011 *Phys. Rev. Lett.* **107** 064301

[10] Lemoult F, Fink M, and Lerosey G 2011 *Waves in Random and Complex Media* **21** 614

[11] Lemoult F, Fink M and Lerosey G 2012 Nat. Commun. **3**, 889

[12] Born M and Wolf E 1980 *Principles of Optics* (Oxford: Pergamon Press)

[13] Liu Z Y, Zhang X X, Mao Y W, Zhu Y Y, Yang Z Y, Chan C T and Sheng P 2000 *Science* **289** 1734

[14] Li J and Chan C T, 2004 *Phys Rev E* **70** 055602

[15] Guenneau S, Movchan A, Pétursson G and Ramakrishna S A 2007 *New J. Phys.* **9** 399

[16] Fang N, Xi D, Xu J, Ambati M, Srituravanich W, Sun C and Zhang X 2006 *Nature Materials* **5** 452

[17] Still T, Cheng W, Retsch M, Sainidou R, Wang J, Jonas U, Stefanou N and Fytas G 2008 *Phys. Rev. Lett.* **100** 194301

[18] Boechler N, Eliason J K, Kumar A, Maznev A A, Nelson K A and Fang N 2013 *Phys. Rev. Lett.* **111** 036103

[19] Lemoult F, Kaina N, Fink M and Lerosey G 2013 *Nature Phys.* **9** 55

[20] Sugimoto N 1992 *J. Fluid Mech.* **244** 55

[21] Sugimoto N and Horioka T 1993 *J. Acoust. Soc. Am.* **97** 1446

[22] Bruneau M 2006 *Fundamentals of Acoustics* (London: ISTE)

[23] Mansfield S M and Kino G S 1990 *Appl. Phys. Lett.* **57** 2615

[24] Ashcroft N W and Mermin N D 1976 *Solid State Physics* (Brooks/Cole)

[25] Belov P A, Hao Y, and Sudhakaran S 2006 *Phys. Rev. B* **73** 33108.